\documentclass[sigconf]{acmart}

\setcopyright{none}
\renewcommand\footnotetextcopyrightpermission[1]{} 

\usepackage{multirow}
\usepackage{tcolorbox}
\usepackage{listings}
\usepackage{xcolor,pifont}

\newcommand{\code}[1]{\texttt{#1}}
\newcommand{\name}[0]{TraceFixer}

\newcommand{\islem}[1]{{\color{blue}#1 --\textbf{Islem}}}
\newcommand{\michael}[1]{{\color{brown}#1 --\textbf{Michael}}}

\newcounter{findingCounter}
\definecolor{halfgray}{gray}{0.35}
\definecolor{deepblue}{rgb}{0,0,0.5}
\definecolor{deepred}{rgb}{0.6,0,0}
\definecolor{deepgreen}{rgb}{0,0.5,0}
\definecolor{highlightorange}{rgb}{0.98, 0.92, 0.8}

\lstdefinelanguage{Python}{
	keywords={if, raise, print, and, or, is, None, not, in, elif, else, =, ., <BUGGY_LINE>, <LINE>, <INITIAL_STATE>, <CONTEXT>, <DESIRED_STATE>},
	ndkeywords={},
	ndkeywordstyle=\color{darkgray}\bfseries,
	identifierstyle=\color{black},
	numberstyle=\color{deepred},
	sensitive=false,
	comment=[l]{//},
	morecomment=[s]{/*}{*/},
	morestring=[b]',
	morestring=[b]"
}

\makeatletter
\newcommand\HL{%
	\gdef\lst@alloverstyle##1{%
		\fboxrule=0pt
		\fboxsep=0pt
		\colorbox{lightgray}{\strut##1}%
	}%
}
\newcommand\HLoff{%
	\xdef\lst@alloverstyle##1{##1}%
}

\begin{document}

\title{\name{}: Execution Trace-Guided Program Repair}

\author{Islem Bouzenia}
\email{fi\textunderscore bouzenia@esi.dz}
\affiliation{%
	\institution{University of Stuttgart}
	\country{Germany}
}

\author{Yangruibo Ding}
\email{yrbding@cs.columbia.edu}
\affiliation{%
	\institution{Columbia University}
	\country{USA}
}

\author{Kexin Pei}
\email{kpei@cs.columbia.edu}
\affiliation{%
	\institution{Columbia University}
	\country{USA}
}

\author{Baishakhi Ray}
\email{rayb@cs.columbia.edu}
\affiliation{%
	\institution{Columbia University}
	\country{USA}
}

\author{Michael Pradel}
\email{michael@binaervarianz.de}
\affiliation{%
	\institution{University of Stuttgart}
	\country{Germany}
}
\begin{abstract}

When debugging unintended program behavior, developers can often identify the point in the execution where the actual behavior diverges from the desired behavior.
For example, a variable may get assigned a wrong value, which then negatively influences the remaining computation.
Once a developer identifies such a divergence, how to fix the code so that it provides the desired behavior?
This paper presents \name{}, a technique for predicting how to edit source code so that it does not diverge from the expected behavior anymore.
The key idea is to train a neural program repair model that not only learns from source code edits but also exploits excerpts of runtime traces.
The input to the model is a partial execution trace of the incorrect code, which can be obtained automatically through code instrumentation, and the correct state that the program should reach at the divergence point, which the user provides, e.g., in an interactive debugger.
Our approach fundamentally differs from current program repair techniques, which share a similar goal but exploit neither execution traces nor information about the desired program state.
We evaluate \name{} on single-line mistakes in Python code.
After training the model on hundreds of thousands of code edits created by a neural model that mimics real-world bugs, we find that exploiting execution traces improves the bug-fixing ability by 13\% to 20\% (depending on the dataset, within the top-10 predictions) compared to a baseline that learns from source code edits only.
Applying \name{} to 20 real-world Python bugs shows that the approach successfully fixes 10 of them.

\end{abstract}


\maketitle
\pagestyle{plain}


\section{Introduction}

When trying to fix a bug,  a developer must localize the problem and then edit the source code to prevent the problem from happening.
During the first step, a developer typically identifies a source code location and understands how the behavior at this location differs from the desired behavior.
For example, when stepping through a program in an interactive debugger,  a developer may realize a point in the execution where the value of some variable differs from the value the variable should have.
We refer to the point during an execution where the actual behavior diverges from the desired behavior as the \emph{divergence point} of a bug, and we call the state the program should reach at this point the \emph{desired state}.
Given these two pieces of information, the second step is to edit the source code in a way that prevents divergence and reaches the desired state.

\begin{figure*}
    \includegraphics[width=\linewidth]{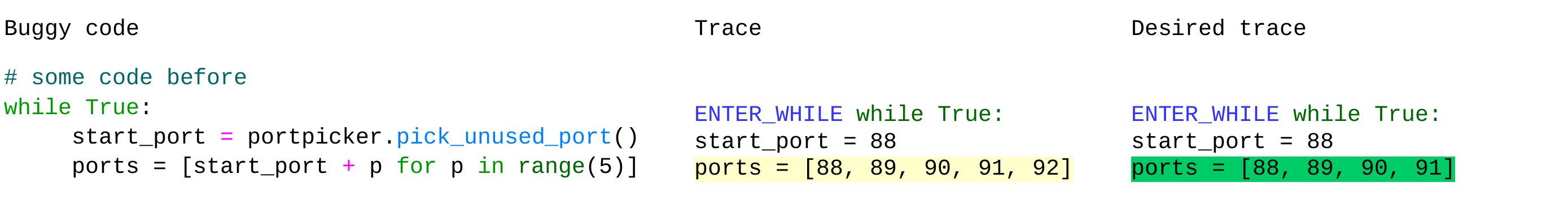}
    \caption{Example of a buggy program with corresponding execution traces.}
    \label{fig:example}
\end{figure*}


For example, Figure~\ref{fig:example} shows a snippet of code taken from the deepmind/pysc2 project\footnote{https://github.com/deepmind/pysc2/commit/803de8a0f707efc3967fde683836c202c73215af}, where the program behavior diverges at the last shown line.
The middle part of the figure gives the trace obtained by executing the buggy code, while the right part shows the trace the developer intends the program to produce.
The difference is in the state produced at the divergence point:
Instead of storing \code{[88, 89, 90, 91, 92]} in variable \code{ports}, the desired state is that \code{ports} has the value \code{[88, 89, 90, 91]}. 

Automated program repair has made impressive progress in fixing programming mistakes over the past few years~\cite{cacm2019-program-repair}, and learning-based approaches are shown to be the current state of the art~\cite{Chen2019,Lutellier2020,ding2020patching,zhu2021syntax}.
However, none of these approaches considers runtime behavior, such as a partial execution trace or the desired state, as an input to the approach.
Instead, many repair approaches validate candidate repairs against test cases of the expected behavior as a post-processing step after predicting fix candidates~\cite{Weimer2009,LeGoues2012,Chen2019,Lutellier2020,Zhu2021} or use tests as feedback for repair predictions~\cite{Ye2022a}.
As illustrated by the example in Figure~\ref{fig:example}, determining the intended fix without knowing what state the program should reach is clearly difficult.
Providing such information to a learning-based repair technique hence offers an opportunity to further improve their effectiveness.


This paper presents \name{}, the first learning-based program repair technique guided by execution traces.
By analogy to the human way of debugging, our approach considers a program's execution trace alongside its code.
The inputs to the approach are the source code of a buggy program, an excerpt of the execution trace produced by this program, and the desired state at the  divergence point.
Given these inputs, the approach predicts a variant of the code that produces the desired trace.
\name{} is enabled by a neural model that reasons about both code and corresponding execution traces to predict the proper fix.

In the example of Figure~\ref{fig:example}, \name{} starts by instrumenting and 
executing the program using, e.g., inputs provided by a user or a test case. 
The result is an execution trace, as shown in the middle part of the figure.
Based on this execution, the developer determines the divergence point and gives the desired state of the program at this point.
For the example, the developer indicates that the last shown line is the divergence point, and that the \code{ports} variable should have the value \code{[88, 89, 90, 91]}.
Given the incorrect code, its execution trace, and the desired state, \name{} predicts how to fix the code by suggesting to replace the last shown line with \code{ports = [start\_port + p for p in range(4)]}.
Indeed, with this fix, the program will produce the desired state.

Creating a trace-guided neural repair approach leads to two key challenges not addressed in prior work.
(C1) First, since our approach is data-driven, we need an extensive dataset that consists of pairs of buggy code and the corresponding correct code, along with execution traces of their behavior.
(C2) Second, we need to identify a suitable neural model and an effective format for feeding the different input modalities--buggy code, execution trace, and desired state--into the model.

\name{} addresses challenge~C1 through a neural bug injection model that creates hundreds of thousands of realistic bugs from code with existing test cases.
Inspired by the success of existing neural approaches to generate bugs~\cite{allamanis2021self,fse2022-vuln,fse2021}, we train a neural bug injection model for Python that mimics single-line bugs observed to occur in the wild~\cite{kamienski2021pysstubs}.
To address challenge~C2, \name{} provides an input formatting module that represents the different modalities in a way that is compatible with popular sequence-to-sequence models, allowing us to build upon a strong pre-trained model~\cite{wang-etal-2021-codet5}.

Our evaluation applies \name{} to single-line bugs in Python.
We inject hundreds of thousands of realistic bugs into two large-scale datasets of executable programs and then train the repair model to fix these bugs.
Once trained, the model successfully repairs 57\%--63\% of the bugs with its top-most suggestion, and 82\%--87\% of the bugs within the top-10 suggestions.
Out of a set of 20 real-world Python bugs, \name{} successfully fixes 10.
Finally, we compare our approach with a neural baseline model that, similar to prior techniques~\cite{Chen2019,Lutellier2020,ding2020patching,zhu2021syntax}, learns only from code but not runtime information, and find that \name{} improves upon the state of the art by 13\% to 20\%.

While practical techniques for finding the divergence point and obtaining the traces provided to \name{} are beyond the scope of this paper, we envision at least two ways of using our approach in practice.
First, we believe that appropriate tool support will make it relatively easy for developers to provide this information when they are anyway localizing a bug.
For example, an interactive debugger that allows developers to step through their code and inspect runtime values could provide an option for marking a specific value as wrong and for providing the correct value instead.
Second, the divergence point and traces could be gathered automatically with existing fault localization techniques~\cite{wong2016survey} applied to test cases that expose the problem, which generate-and-validate repair techniques~\cite{cacm2019-program-repair} commonly assume to exist.

In summary, this paper contributes the following:
\begin{itemize}
	\item An automated repair technique that learns not only from source code but also from runtime traces and information about the state a program should reach.
	\item A neural repair model enabled by learning from a large-scale set of automatically injected bugs and a novel input formatting that feeds different modalities into a sequence-to-sequence architecture.
	\item Empirical evidence that considering execution traces improves the bug-fixing abilities of the approach and that \name{} is effective for real-world Python bugs.
\end{itemize}

\section{Approach}

In this section, we present our data-driven, deep learning-based automated program repair approach called \name{}.
We start by defining the problem we address (Section~\ref{sec:pb}), then give an overview of the approach (Section~\ref{sec:overview}), and finally explain its different components in detail (Sections~\ref{sec:injection} to~\ref{sec:learning}). We also give a preliminary background in Section~\ref{sec:preli}

\subsection{Problem Definition}
\label{sec:pb}

\name{} reasons about executions of programs, which we represent as a sequence of states:

\begin{definition}[State] 
	\label{def:state}
	The state $s$ at a point during the execution of a program $p$ is a list of pairs $(n, v)$ where the $n$ is the name of a variable available in the current scope during the execution of $p$ and $v$ is the value that $n$ refers to.
\end{definition}

We consider both primitive values, e.g., integers and strings, and non-primitive values, e.g., a list or an instance of a class.
The approach focuses on the state in the program's main memory and ignores any other state, such as the state of the file system, the state reachable via the network, or the state of another process.


During execution, a program transfers from one state to another.
We represent these transfers in an execution trace:

\begin{definition}[Execution trace] 
	\label{def:trace}
	The execution trace of program $p$ is a pair $(s_{init}, E)$, where $s_{init}$ is the initial state and $E$ is a sequence of events observed starting from the initial state.
	Each event in $E$ is a pair $(l_i, s_i)$ where $l_i$ is an executed line of code in $p$, and $s_i$ is the state after executing the line.
\end{definition}

The full execution trace until triggering a bug may consist of many more events than a developer would typically inspect when reasoning about an incorrect execution.
Instead of considering the full execution trace starting at the first executed line of the program, the initial state in a trace may be any point in the program's execution.
In our evaluation, we consider traces with up to three events.
If the line $l_i$ captured in an execution trace involves a function call, then we include the return value of the call (if any) as part of the state $s_i$.

The usage scenario of \name{} is that a developer identifies the point during an execution where the actual and the intended execution start to diverge:

\begin{definition}[Divergence point]
	\label{def:divergence}
	Consider two traces $(s_{init}, E)$ and $(s'_{init}, E')$ with
	$E=[(l_1, s_1), ..., (l_k, s_k)]$,
	$E'=[(l'_1, s'_1), ..., (l'_k, s'_k)]$,
	$l_i = l'_i$ and $s_i=s'_i$ for all $i<k$,
	but $l_k \neq l'_k$ or $s_k \neq s'_k$.
	That is, the traces are equal except for the last event, where either the line that gets executed, what state is reached after executing the line, or possibly both differ.
	The last event in the two traces is called the divergence point.
\end{definition}

Once a developer has identified a divergence point, e.g., while stepping through a program using an interactive debugger, the problem addressed in this paper is how to fix the incorrect code:

\begin{definition}[Trace-guided repair problem] 
	\label{def:problem}
	Given a buggy piece of code $c$, a trace $t$ produced by this code, and a state $s$ that is the desired state of the program at the divergence point, the problem is to predict the fixed code $c'$ that, if executed, will yield a trace that differs from $t$ only by reaching state~$s$.
\end{definition}

\subsection{Overview}
\label{sec:overview}

\begin{figure}
	\includegraphics[width=0.9\linewidth]{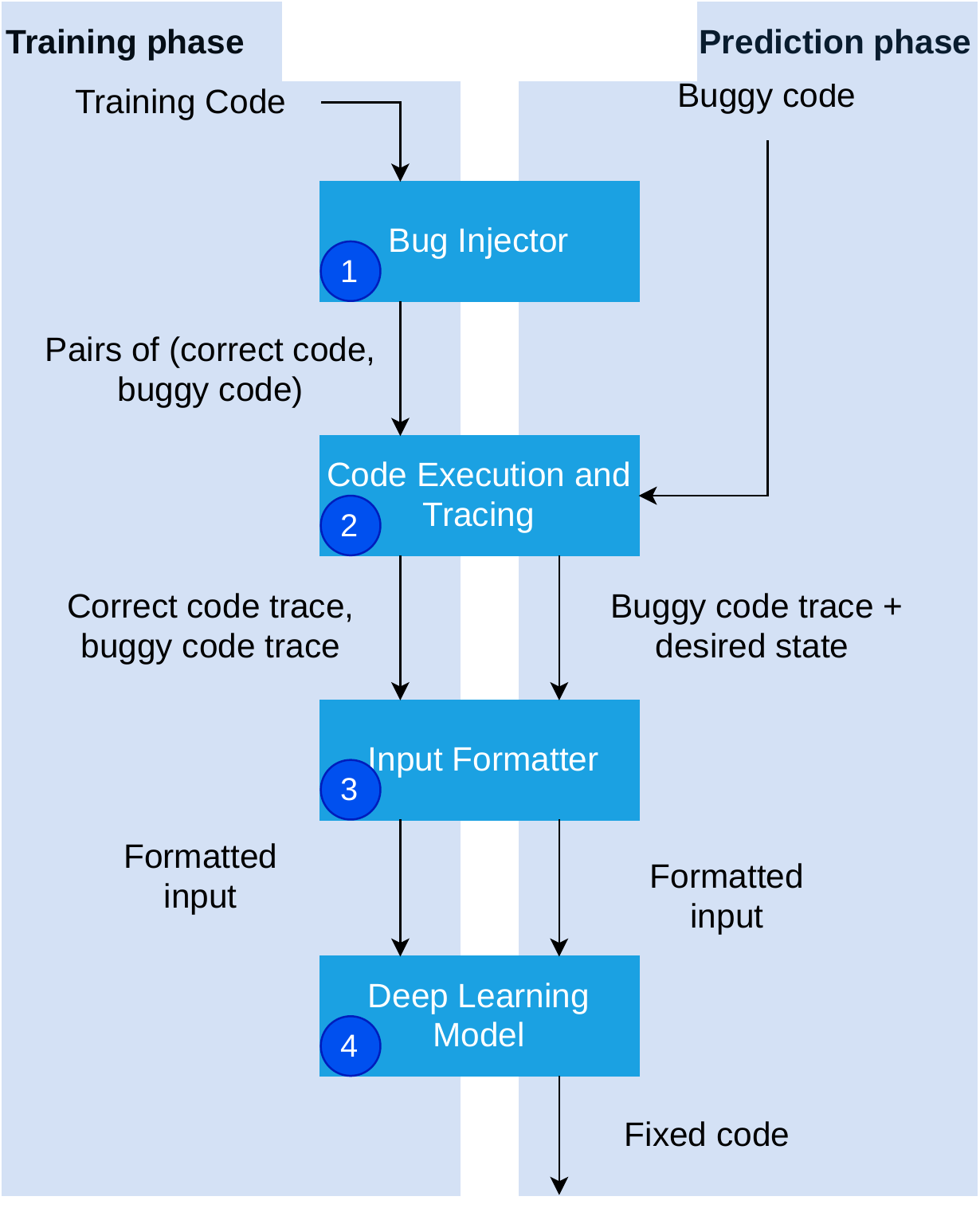}
	\caption{Overview of TraceFixer approach.}
	\label{fig:approach_overview}
\end{figure}

Our approach to addressing the abovementioned problem consists of four main steps represented in Figure~\ref{fig:approach_overview}. 
(i) \textit{Bug Injector.} The first step is specific to the training phase. In every deep learning-based application, having a training dataset is crucial. In our case, we first need a dataset composed of pairs of buggy code and the corresponding correct code. 
Therefore, we design a neural bug injector: given a correct code and a location, this step injects bugs by modifying the code at the given location. We leverage a pre-trained seq2seq model, which receives a correct program with the injection location (target line) as input and outputs the buggy version of the target line. At a high level, this step can be thought of using a seq2seq model in the reverse manner of neural bug fixing~\cite{chakraborty2021modit}. The injector further replaces the target line in the correct program with the generated buggy line.
(ii) \textit{Code Execution and Tracing.}  Once we have pairs of correct and buggy code, we instrument and execute them to get their execution traces. In the training phase, the approach extracts the divergence point and the desired state from these traces. In the prediction phase, the developer directly provides this information.
(iii) \textit{Input Formatter.}
The buggy code, the trace of the buggy code, and the desired state will be passed into a neural model.
To this end, we format the different data modalities in a way suitable for seq2seq architectures. 
(iv) \textit{Deep Learning Model.} In the final step, we train a neural model to fix the buggy code and, once trained, use the model to predict the fix for previously unseen bugs.

\subsection{Preliminary}
\label{sec:preli}
\paragraph{\textbf{Transformer-based sequence-to-sequence model}} 
In recent years, Transformer based sequence-to-sequence (seq2seq) models have been applied to code editing tasks~\cite{chakraborty2021modit, zhang2022coditt5, ding2020patching, Ahmad2021, wang-etal-2021-codet5}. The model contains an encoder and a decoder, and both are built based on a stack of Transformer layers~\cite{Vaswani2017}. The encoder summarizes the information of the input sequence, and a decoder generates the new token sequences based on the output of the encoder. The decoder predicts the new sequence left-to-right, token by token. Every token is generated conditioned on both the encoded input and the generated prefix (i.e., left tokens already generated in the previous time steps). The model's input is the flattened sequence of a program, and the expected output is the edited code so that the model will learn how to change the input to match the ground truth during training.

\paragraph{\textbf{CodeT5}} CodeT5~\cite{wang-etal-2021-codet5} is a state-of-the-art pre-trained seq2seq model for source code. The pretraining is done across different code-generation tasks, including code summarization, code generation, and code refinement. The code refinement task is the most related task to our work: it randomly masks code tokens of the input with a special token \texttt{<MASK>}, and trains the model to predict the masked tokens back. Conceptually, this task teaches the model to edit the code, which can potentially assist with our proposed models to inject or fix bugs.
CodeT5 is pre-trained with millions of samples with more than 100 epochs. With such an effective pretraining, it reports promising results in both code understanding and generation tasks. In addition, CodeT5 provides two model sizes in the original paper: small and base. The former has six layers of transformers for both encoder and decoder, while the latter has 12 layers for both encoder and decoder.

\paragraph{\textbf{Byte-pair Encoding}} The vocabulary for source code is essentially open-ended since developers could name identifiers with any preferred combination of words or introduce new words that are not part of any natural language's vocabulary~\cite{Karampatsis2020a}. To train a neural network on source code, however, we need a fixed length of vocabulary~\cite{hellendoorn2017are} while minimizing the out-of-vocabulary cases. Byte-pair encoding (BPE) was initially applied to address the rare words issue in neural machine translation~\cite{Sennrich2016}: the algorithm splits the rare and unknown words into sub-words already part of the vocabulary; in the worst case, a complicated word will be split into characters. Similarly, we apply BPE to address the out-of-vocabulary issue of source code in our models: the common code tokens, such as keywords, will be encoded as single tokens, while the complicated code tokens will be split into common sub-tokens in the vocabulary.

\begin{figure*}[!ht]
    \centering
    \includegraphics[width=\linewidth]{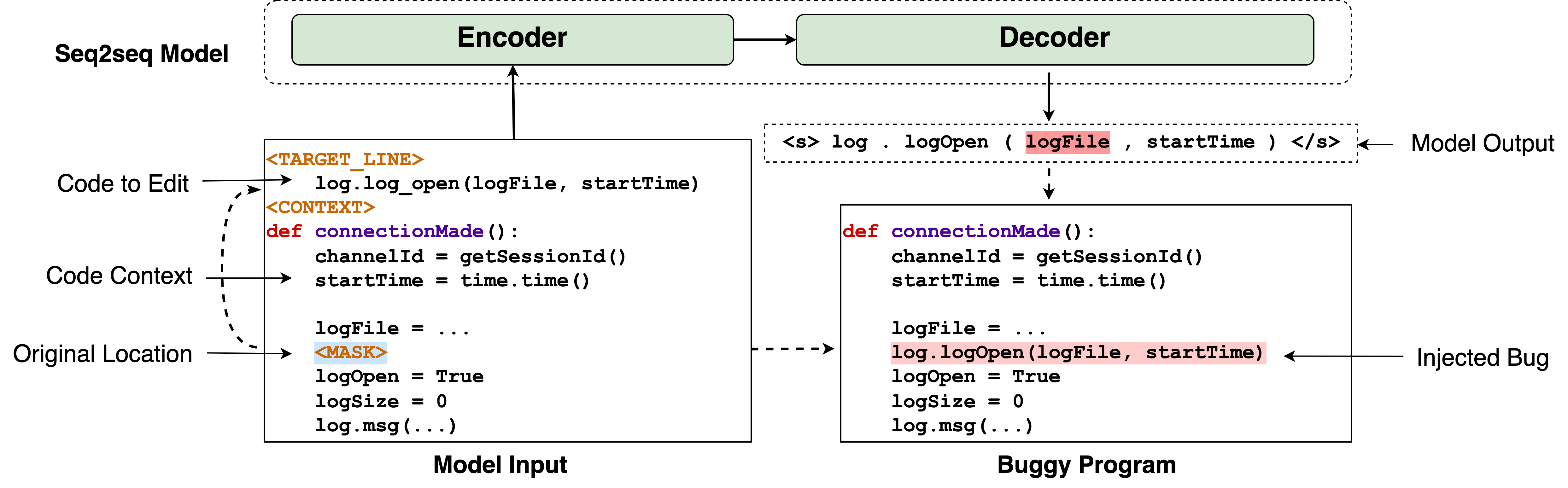}
    \caption{Workflow of bug injection.}
    \label{fig:bug_inject}
\end{figure*}

\subsection{Bug Injection}
\label{sec:injection}

To train a neural model to repair bugs, first, we need a dataset composed of the buggy and correct code pairs. The dataset should be large enough to enable effective learning by a neural model. 
Each buggy and correct code pair in the dataset should be executable to allow gathering traces. Last but not least,  the injected bugs should resemble real-world bugs. To this end, we use a neural bug injection technique, motivated by the success of existing neural approaches to generate realistic bugs~\cite{allamanis2021self,fse2022-vuln,fse2021}.

The backbone of our tool is a Transformer-based, sequence-to-sequence (seq2seq) neural network. To save the training efforts and ensure the code generation quality, we initialize our neural network with CodeT5-small~\cite{wang-etal-2021-codet5}, which has been trained on millions of code samples with a code refinement task, so it already has good knowledge about how to edit code. Consequently, loading it as the initial model can help our neural bug injector converge faster during training and generate more valid programs than random initialization.

We use the PySStuBs dataset~\cite{kamienski2021pysstubs} to train the neural bug injector. This dataset contains samples with both buggy and patched versions, and we build the model input with the patched version while the ground truth is the corresponding bug. Consequently, we train the neural model to inject bugs by reversing the bug-fixing process. Furthermore, we ignore the samples that cannot be checked out due to the invalid commit or file hashes, and we do not consider the files written with Python2. We end up with 47,314 valid samples; we split them into 95\% and 5\% portions for training and tuning parameters, respectively.

The workflow of the bug injector is shown in Figure~\ref{fig:bug_inject}. The first step of the tool is to create the input for the neural model based on a correct program and a target line at which to inject bugs. The correct program will be a patch in the PySStuBs dataset, and the target line is the changed line when fixing the corresponding bug. The bottom-left of Figure~\ref{fig:bug_inject} shows an example of the model input, where \texttt{log.log\_open(logFile, startTime)} is the target line. To ensure the model performance, we design the model input following the state-of-the-art neural code editor, \textsc{Modit}\cite{chakraborty2021modit}, where we extract the line to be edited from the program and prepend it at the beginning of the input sequence. We use special tokens to split the target line and its contexts so the model will know their positions. To further indicate the original location of the target line and align with the design of the pretrained CodeT5, we fill that location with a placeholder token, \texttt{<MASK>}.

Given the model input, the second step is to train the seq2seq model to predict the buggy version of the target line. During training, the model will be fed with a ground-truth bug, and the model learns to construct the bug left-to-right, token-by-token. Later, we replace the \texttt{<MASK>} in the input with the model prediction to generate the buggy program.


\subsection{Execution and Tracing}
\label{sec:executing}



Since our approach is based on program repair using traces, getting those traces is essential and required in both the training and the prediction phase. For that, we instrument the target code and then execute it. Our instrumentation targets the line level, meaning that we collect the program state after executing every line of code which matches definition \ref{def:trace}. However, we limit the instrumentation and the collected trace to a specific scope. In our case, the scope is usually one function or a sequential script consisting of one file. We intentionally do not trace subsequent calls to other functions or modules, mimicking a developer who steps through the execution of a function without stepping into callees. In the case of a return statement, we copy the value of the returned variable before it is returned, and we copy the new state if the return expression changes it.

When collecting the traces, we save the printable string of all variables in scope in a format similar to the one of definition \ref{def:state}. For example, if at some point of execution we have \code{primes = [2, 3, 5, 7, 11]}, the saved state related to the variable \code{primes} would be \code{[("primes", "[2, 3, 5, 7, 11]")]}. Saving the evaluation of the variable "primes" as a string is easy because the printable string of the variable primes matches the value that a developer cares about, which is, in this case, the list of numbers themselves. Unfortunately, not all variables are printable like primitive types and standard data structures. For that, if the value of a variable is not a primitive type or a standard data structure, we inspect the object referred to by the variable. We collect the values of all printable attributes of that object. For example, after this line of code: \code{rectangle = Rectangle(2, 5)}, the variable \code{rectangle} will refer to an instance of the class \code{Rectangle}, which has two attributes: \code{width} and \code{height}. The state of the program after that line would be \code{[("rectangle.width", "2"), ("rectangle.height", "5")]}. In the case where there are no printable values when inspecting the attributes of an object, we show the type of the variable. For example, if the variable \code{rectangle} has no printable attributes then the trace of the line \code{rectangle=Rectangle(2, 5)} would be \code{[("rectangle", "instance(Rectangle)")]}.

In the following, we describe the flow of instrumentation and execution to get the traces. 
\paragraph{\textbf{Training phase}} We need the traces of both buggy programs and their corresponding correct version for the training. Given a set of pairs \code{(buggy program $ p $, fixed version $ p' $)}, we first start by filtering out syntactically incorrect code. Next, we instrument the code of both $ p $ and $ p' $to get the traces on a line level as defined in Definition~\ref{def:trace}. Then, we execute the instrumented programs while automatically excluding programs terminated with a runtime exception as it is out of the scope of our approach since we are interested in divergence points caused by the difference in the value of the program's state as defined in Definition~\ref{def:state}. Finally, we compare the trace of each successfully executed buggy program to the trace of its correct version to keep only the buggy programs having a trace that has a divergence point from the trace of the correct version. In some programs, the traces are non-deterministic, e.g., due to the existence of a variable or expression that depends on a random generator or a variant factor (like time). To determine those non-deterministic traces, we execute the original correct code (of the program $ p' $) using the same test case at two different timestamps. If the two executions produce different traces, we consider it non-deterministic and drop it. We drop non-deterministic instances because the trace between the original and buggy codes might differ because of that random statement, even if it is not the bug's location. After getting the correct and buggy code trace, we determine the point of divergence by doing a diff between the two traces (correct and buggy). The first location of difference is considered to be the point of divergence. The program's trace ends at the determined point of divergence, and any collected trace after that point is removed. At the end of this step, we get a list of data
points \code{(buggy trace $t$, desired state $s$, buggy code $ c $, correct code $ c' $)}. To get the desired state, we first locate the point of divergence and copy the state of the correct program at the point of divergence, which becomes the desired state.


\paragraph{\textbf{Prediction phase}}  In this phase, the developer executes their code and gets the traces. Then, upon inspecting the trace, the developer locates the divergence point and gives the corresponding desired state. At the end of this step, we get a prediction data point \code{(buggy trace $t$, desired state $s$, buggy code $ c $)}.

\label{subsec:instrumentation}

\subsection{Constructing and Formatting Input}



After getting the traces of the programs, we construct the input data for the neural model. The input formatting module takes a data point and then transforms it into a format compatible with the model's input layer.
Similar to most previous work on program repair~\cite{cacm2019-program-repair}, we assume that the location where the code needs to be fixed is known, as this location can be obtained via existing techniques~\cite{wong2016survey}.

\begin{definition}[Formatted data point] 
	\label{def:formatted}
	a formatted training data point is a pair \code{(source, target)} where the \code{source} is the input given to the model, and the \code{target} is the ground truth for what the model should predict. The input \code{source} and \code{target} are defined as follows: 
	\begin{itemize}
		\item Source: concatenation of the following modalities separated by the special tokens ($ <..> $). \begin{itemize}
			\item <BUGGY\_LINE> the buggy target line $ l_{b} $
			\item <INITIAL\_STATE> initial state $s_{init}$ of the program $ p $ 
			\item <LINE> executed line of code <STATE> the programs state
			\item <DESIRED\_STATE> the desired or expected correct state
			\item <CONTEXT> source code of the program
		\end{itemize}
	\item Target: \begin{itemize}
		\item <START> the fix of the given buggy line <END>
	\end{itemize}
	\end{itemize}

In prediction phase, a data point consistent of \code{source} only. We construct the formatted data point from elements in the data point resulting from step two.
\end{definition}



Figure~\ref{fig:formatexample} shows an example of the input format that follows the format of definition~\ref{def:formatted}, including the different modalities. The special separators \code{<..>} inserted between the modalities are in bold text. The red dashed line surrounds the buggy line, while the green dashed line delimits the program's state, followed by the desired state below (blue dashed lines). We also provide the context as the last modality. In the example of Figure~\ref{fig:formatexample}, we represent and give the model the last three states of the buggy code trace. The trace size in terms of the number of states is a configurable parameter in our approach, and it represents the number of traced lines before the divergence point (included). Furthermore, we only include variables that get changed from one state to another. For example, we did not mention the variable \code{start\_port} in the last state because there was no change in its value from the second state to the last one. All the input modalities together form the \code{source} input. The other part is the \code{target} delimited by the start and end tokens. The target is the ground truth used to evaluate the model's prediction. 

\subsection{Deep Learning Model}

\begin{figure*}
	\includegraphics[width=0.95\linewidth]{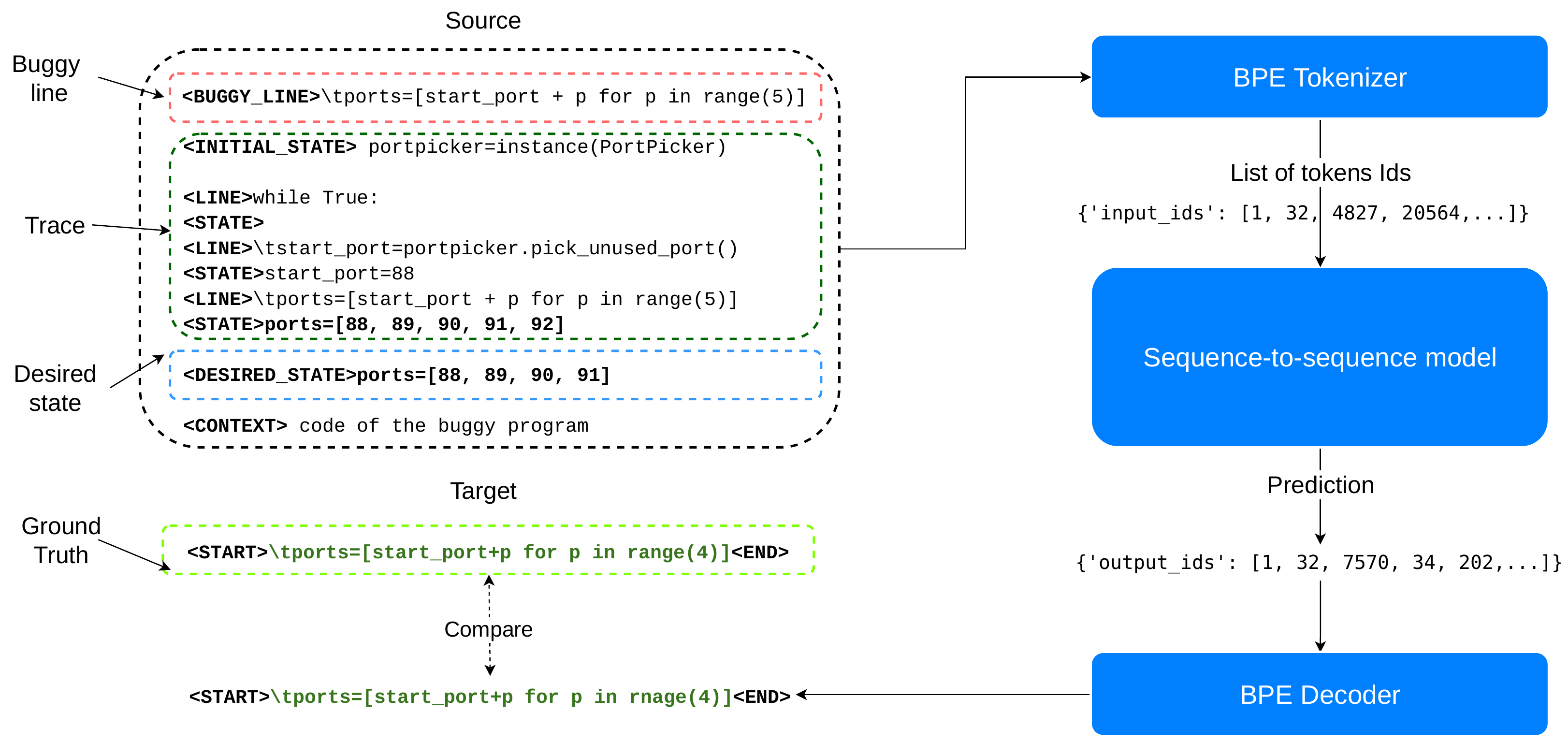}
	\caption{Example of input/output format of the neural model (\code{<..>} denote special tokens).  }
	\label{fig:formatexample}
\end{figure*}
\label{sec:learning}

Once we format the input, we tokenize it using BPE tokenizer (explained in section \ref{sec:preli}). The BPE tokenizer has an encoding function that encodes a string into a list of tokens Ids. A token Id is a number that uniquely encodes and designates a token in the vocabulary of the BPE tokenizer. Figure~\ref{fig:formatexample} shows the first tokens Ids resulting from encoding the source input through BPE tokenizer. The BPE tokenizer also has a decoding function that decodes a list of tokens Ids into a string, as shown in the example of Figure~\ref{fig:formatexample}.

After tokenization and encoding, the final step in our approach is training the neural model. Our model is a sequence-to-sequence Transformer based on CodeT5, explained in section \ref{sec:preli}. At each training step, the model receives a batch of data points, predicts the corresponding fixes and gets evaluated against the ground truth targets by calculating the loss and then updating the weights accordingly. Given a prediction data point, the model outputs a sequence of tokens Ids, which the BPE then decodes to construct a string corresponding to the predicted fix.

\section{Evaluation}

To evaluate the approach, we address the following research questions:
\begin{itemize}
	\item RQ1: How effective is \name{} at fixing bugs?
	\item RQ2: How does \name{} compare to a baseline approach that takes only the code, but no traces, as its input?
	\item RQ3: How effective is a multi-task model trained to handle both trace-guided repair and code-only repair?
	\item RQ4: How does the effectiveness of \name{} vary across different types of bugs?
	
\end{itemize}

\subsection{Experimental Setup}

\subsubsection{Datasets}

We use three existing datasets that offer real-world code and bugs in Python.

\paragraph{\textbf{PySStuBs}} This dataset~\cite{kamienski2021pysstubs} is an extensive collection of single statement bugs in popular open-source Python projects.
We use this dataset in two ways.
First, we train the bug injector described in Section~\ref{sec:injection} on the PySStuBs bugs.
Second, we sample 20 bugs from PySStuBs to assess whether \name{} can fix real-world bugs.
The reason for focusing on 20 bugs is that we have to manually set up executable tests that trigger each bug.
The setup is non-trivial as each project in PySStuBs has its own requirements and many are no longer compatible with the latest versions of packages and even Python.  We made sure these 20 bugs do not overlap with the training dataset. 



\paragraph{\textbf{CodeNet}} This dataset~\cite{DBLP:conf/nips/Puri0JZDZD0CDTB21} is a set of algorithmic challenges where participants try to solve a particular problem and submit their code to the platform to get automatically evaluated against test cases.
We use 3,200 problems, where we take three submissions from each problem, adding up to 9,600 different programs.
Since the CodeNet dataset does not provide buggy versions of the programs, we automatically inject bugs (Section~\ref{sec:injection}).
For each program, the bug injector injects three different bugs at every line of code, which yields 850K buggy programs.
Following the steps described in Section~\ref{sec:executing} we end up with 103K data points.
We keep 500 of them for testing and use the rest to train the neural model.

\paragraph{\textbf{TheAlgorithms}} This dataset\footnote{\url{https://github.com/TheAlgorithms/Python}} is a collection of programs that solve algorithmic problems, along with test cases to validate their correctness.
The problems cover various domains, such as arithmetic and algebra, neural networks, and web programming.
Similar to the CodeNet dataset, we inject bugs into the given programs with a rate of five variants per line, which yields 120K buggy programs.
From the resulting 8.5K datapoints, we keep 300 for testing and use the rest for training.

\subsubsection{Metrics}

We measure the effectiveness of~\name{} by checking if the suggested fixes match the known ground-truth fixes.
A suggested fix counts as correct if and only if it exactly matches the ground truth,  except for unnecessary whitespace.
Similar to prior work on learning-based repair~\cite{Chen2019,Lutellier2020}, we query the model to produce multiple fix suggestions using beam search.
We report how often the model finds the correct fix among the unique top-k suggestions, or short UTOPk.
The UTOPk metric is a variant of the usual top-k accuracy that ignores duplicate predictions and keeps querying the model until we find k unique fix suggestions.
To determine duplicates, we first start by removing unnecessary tokens from the predicted line of code. For example, \code{a = a + 1} and \code{a=a+1} are duplicates because the only difference is unnecessary whitespace.

\subsubsection{Baseline}

The key contribution of this work is to introduce execution traces into learning-based repair.
To assess the benefits of this idea, we compare \name{} to a baseline model that takes only the buggy source code as its input, called \emph{Code-only}.
Conceptually, this model follows the source code-only, sequence-to-sequence approach proposed in recent learning-based repair techniques~\cite{Chen2019,Lutellier2020, chakraborty2021modit}. 
We implement the baseline as a stripped-down variant of \name{}, which differs from the full \name{} approach only by not receiving any execution traces as the input.  This code-only setting is similar to prior work,  MODIT~\cite{chakraborty2021modit}---both use similar input formatting. 
A possible alternative would be to compare to the implementation of prior code-only repair techniques directly.
We decided against that setup (i) because prior tools focus on Java instead of Python and (ii) because it would bias the results due to details of the techniques that are not directly related to the question of whether traces are helpful for learning-based repair.

\subsubsection{Implementation and Hardware}

We implement the different components of \name{} into a Python-based tool. To instrument and collect traces, we use Python standard modules such as SetTrace and Inspect alongside other community packages like beeprint.\footnote{https://github.com/panyanyany/beeprint} The neural models are based on CodeT5-small shared through hugging face api.\footnote{https://huggingface.co/Salesforce/codet5-small}. 
We conduct our experiments on two GPU machines. The first machine is equipped with an NVIDIA Tesla P100 (16GB), while the second has an NVIDIA Tesla V100 (32GB).
To save time, we split our experiments across the two machines, which allows us to train one epoch in about 2.5 hours on the first machine and in about 1.5 hours on the second machine.

\subsection{RQ1: Bug Fixing Effectiveness}
\label{subsec:RQ1}



\begin{table*}[]
	\caption{UTOPk accuracy on different datasets.}
	
	\begin{tabular}{clrrrr}
		
		\hline
		\multicolumn{1}{l}{Dataset} & Models & \multicolumn{1}{l}{UTOP1(\%)} & \multicolumn{1}{l}{UTOP3(\%)} & \multicolumn{1}{l}{UTOP5(\%)} & \multicolumn{1}{l}{UTOP10(\%)} \\ 
		\hline

		\multirow{2}{*}{CodeNet}       & \name{}  & \textbf{63} & \textbf{78} & \textbf{82} & \textbf{87} \\
		& Code-only & 50 & 65 & 69 & 78 \\
		\hline
		
		\multirow{2}{*}{TheAlgorithms} & \name{}  & \textbf{57} & \textbf{75} & \textbf{77} & \textbf{82} \\
		& Code-only & 49 & 64 & 66 & 69 \\ 
		\hline
		
		\multirow{2}{*}{20 Real bugs} & \name{}  & 10 & \textbf{25} & \textbf{30} & \textbf{50} \\
		& Code-only & 10 & 15 & 25 & 30 \\ 
		\hline
	\end{tabular}
	\label{tab:tabrq1}
\end{table*}

To investigate the effectiveness of \name{} in program repair, we evaluate our approach on three different datasets: CodeNet, TheAlgorithms and 20 real bugs sampled from PySStuBs. For the CodeNet and TheAlgorithms datasets, we finetune a separate CodeT5-small model for each dataset and then evaluate on a held-out test set from the same dataset. To evaluate \name{} on the 20 real bugs, we use the model trained on CodeNet.

The results of the evaluation on the test sets are shown in Table~\ref{tab:tabrq1}.
The table shows that \name{} achieves high accuracy across the three setups.
For example, considering only the top-most prediction by the model fixes 63\%, 57\%, and 10\% of the bugs, while the predictions in the UTOP10 contain correct fixes for 87\%, 82\%, and 50\% of all bugs.
The fixing accuracy is the lowest on the set of real bugs, which we attribute to three reasons.
First, the model is trained on a different dataset (CodeNet) and then applied to the real-world bugs, which may cause a shift of distribution~\cite{He2022}.
Second, the real bugs are extracted from open-source projects that are more complicated than the solutions to algorithmic problems in the CodeNet and TheAlgorithms datasets.
Finally, the real-world bugs occur in code that calls other modules and functions within the same project or other packages. In contrast, the CodeNet code mainly uses standard Python functions and data structures. 

For a more detailed look into the 20 real bugs, Table~\ref{tab:realworld found} shows all of them alongside the fixes and the effectiveness of the model.
We can see that the~\name{} is able to fix different kinds of problems, including control flow-related bugs (e.g., example~4), data structure-related bugs (e.g., examples~8 and~10), and various kinds of incorrect expressions (e.g., examples~1 and~14).
The cases in which~\name{} cannot find a fix often are due to the out-of-vocabulary problem~\cite{Karampatsis2020a}, such as in examples~2, 3 and~18. Out-of-vocabulary here means that the fix requires some tokens that do appear neither in the provided code nor the execution trace, and unless the model has seen such a fix during training, it typically cannot predict the desired tokens. 

\begin{tcolorbox}[colback=blue!5!white,colframe=blue!5!white,arc=0mm,grow to left by=0mm,left=0mm,grow to right by=0mm,left=1.5mm,right=1.5mm,top=1.5mm,bottom=1.5mm]
	\textbf{Finding \arabic{findingCounter}\stepcounter{findingCounter}:}
	Throughout the different evaluation setups, \name{} effectively predicts the correct fix among the UTOP10 in 50\%--87\% of the cases.
\end{tcolorbox}

\subsection{RQ2: Comparison with Code-Only Approach}

\begin{figure}
	\includegraphics[width=\linewidth]{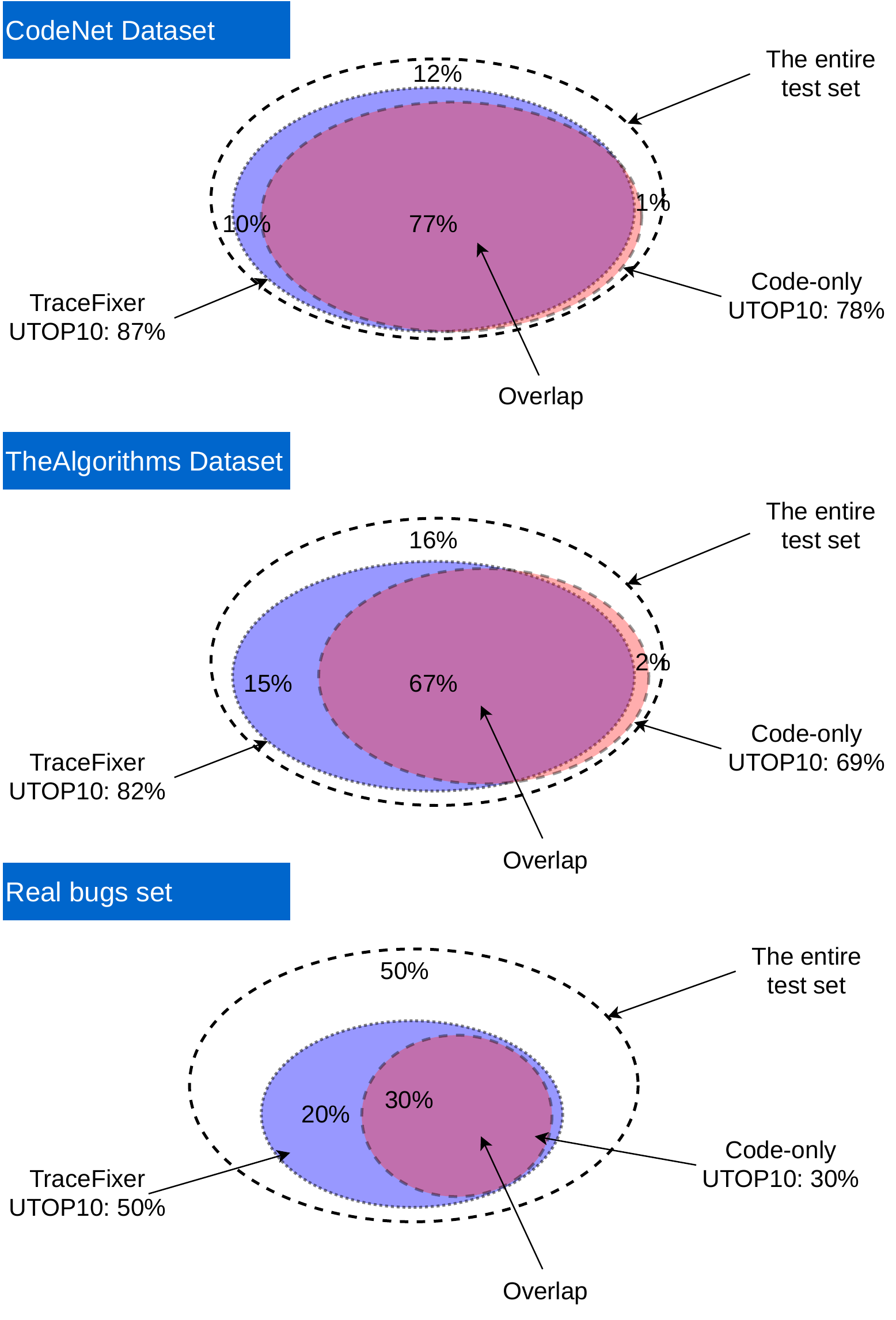}
	\caption{Overlap between \name{} and code-only fixed instances across the three datasets (Venn diagram).}
	\label{fig:venn_diagram}
\end{figure}

The key contribution of \name{} is to use an execution trace and the desired state as an additional input to a repair model. To assess the impact of this contribution over prior work, we compare our approach to the code-only baseline. Table~\ref{tab:tabrq1} shows that across the three evaluation setups, \name{} improves over the code-only baselines in terms of UTOP K accuracy. For example, in the CodeNet evaluation setup, \name{} improves with 13\% over the UTOP 1. Using three different datasets suggests that the improvement over the code-only approach is independent of the dataset. 

Being powered by the trace, our approach provides the model with more information to fix the bug, which aligns with the analogy of the developer's way of debugging. While the code-only model still performs well on the three evaluation datasets, it has limited knowledge about the desired change and thus does not predict the most likely fix in many cases. Our approach mitigates the limitation of code-only models by adding traces and the desired states as data modalities, thus giving more guidance to perform the repair task.

Figure~\ref{fig:venn_diagram} shows the Venn diagram of the overlap in the correctly fixed instances between \name{} and code-only across the three datasets. For example, the first part of Figure~\ref{fig:venn_diagram} shows that \name{} predicts correct fixes among the UTOP10 for almost all the instances that code-only can fix with UTOP10 predictions. Similarly, only 2\% of the TheAlgorithms test set is fixed correctly by the code-only model and not fixed by \name{}. Finally, in the real bugs, the set of fixed instances by the code-only model is strictly included in the set of instances fixed by \name{}. In summary, our approach fixes more than 97\% of the instances that the code-only baseline can fix across the three datasets. Of course, \name{} fixes other instances that the code-only model cannot fix.
\begin{table*}
\caption{ The list of 20 real bugs used in RQ2.}
\label{tab:realworld found}
\setlength{\tabcolsep}{4pt}
\renewcommand{\arraystretch}{0}
\begin{tabular}{@{}rp{20em}p{22em}p{7em}p{3em}@{}}
\toprule
Id & Buggy line & Ground truth and Model's prediction & Project & Correct? \\
\midrule

1 &
\begin{lstlisting}[aboveskip=-.8em,belowskip=-.7em]
num_lines = size_log_area (@\HL@)/(@\HLoff@) 2
\end{lstlisting}
&
\begin{lstlisting}[aboveskip=-.8em,belowskip=-.7em]
GT/Model: num_lines = size_log_area (@\HL@)//(@\HLoff@) 2
\end{lstlisting}
& 
MycroftAI
&
Y
\\ \midrule
2 &
\begin{lstlisting}[aboveskip=-.8em,belowskip=-.7em]
return (@\HL@)((x + y) == 1) * 1.0(@\HLoff@)
\end{lstlisting}
&
\begin{lstlisting}[aboveskip=-.8em,belowskip=-.7em]
GT: return (@\HL@)1 - abs(x + y - 1)(@\HLoff@)
Model: return ((x + y) == 0) * 1.0
\end{lstlisting}
& 
keras
&
N
\\  \midrule
3 &
\begin{lstlisting}[aboveskip=-.8em,belowskip=-.7em]
return rs.rand(3)
\end{lstlisting}
&
\begin{lstlisting}[aboveskip=-.8em,belowskip=-.7em]
GT: return rs.rand(3)(@\HL@).tolist()(@\HLoff@)
Model: return int(rs.rand(3))
\end{lstlisting}
& 
pandas
&
N
\\  \midrule
4 &
\begin{lstlisting}[aboveskip=-.8em,belowskip=-.7em]
if not columns or len(columns) == (@\HL@)9(@\HLoff@):
\end{lstlisting}
&
\begin{lstlisting}[aboveskip=-.8em,belowskip=-.7em]
GT/Model: if not columns or len(columns) == (@\HL@)0(@\HLoff@):
\end{lstlisting}
& 
great-expectations
&
Y
\\  \midrule

5 &
\begin{lstlisting}[aboveskip=-.8em,belowskip=-.7em]
schedule_json["interval"] = query.old_schedule
\end{lstlisting}
&
\begin{lstlisting}[aboveskip=-.8em,belowskip=-.7em]
GT: schedule_json["interval"] = (@\HL@)int((@\HLoff@)query.old_schedule(@\HL@))(@\HLoff@)
Model: schedule_json["interval"] = query.new_schedule 
\end{lstlisting}
& 
redash
&
N
\\  \midrule

6 &
\begin{lstlisting}[aboveskip=-.8em,belowskip=-.7em]
step_size = step_size_scaling (@\HL@)*(@\HLoff@) n**(1/4.)
\end{lstlisting}
&
\begin{lstlisting}[aboveskip=-.8em,belowskip=-.7em]
GT: step_size = step_size_scaling (@\HL@)/(@\HLoff@) n**(1/4.)
Model: step_size = step_size_scaling * n**(1/2.) 
\end{lstlisting}
& 
pymc
&
N
\\  \midrule

7 &
\begin{lstlisting}[aboveskip=-.8em,belowskip=-.7em]
limit = parse_integer(request, "limit", (@\HL@)100(@\HLoff@))
\end{lstlisting}
&
\begin{lstlisting}[aboveskip=-.8em,belowskip=-.7em]
GT/Model: limit = parse_integer(request, "limit", (@\HL@)0(@\HLoff@))
\end{lstlisting}
& 
matrix-org/synapse
&
Y
\\  \midrule

8 &
\begin{lstlisting}[aboveskip=-.8em,belowskip=-.7em]
ports = [start_port + p for p in range((@\HL@)5(@\HLoff@))]
\end{lstlisting}
&
\begin{lstlisting}[aboveskip=-.8em,belowskip=-.7em]
GT/Model: ports = [start_port + p for p in range((@\HL@)4(@\HLoff@))]
\end{lstlisting}
& 
deepmind/pysc2
&
Y
\\  \midrule

9 &
\begin{lstlisting}[aboveskip=-.8em,belowskip=-.7em]
self._ports = _pick_unused_ports((@\HL@)1 + (@\HLoff@)self._num_players * 2)
\end{lstlisting}
&
\begin{lstlisting}[aboveskip=-.8em,belowskip=-.7em]
GT/Model: self._ports = _pick_unused_ports(self._num_players * 2)
\end{lstlisting}
& 
deepmind/pysc2
&
Y
\\  \midrule

10 &
\begin{lstlisting}[aboveskip=-.8em,belowskip=-.7em]
_transfer_block(src, dst.res4, ['4a'] + ['4b%d' % i for i in range(1, (@\HL@)23(@\HLoff@))])
\end{lstlisting}
&
\begin{lstlisting}[aboveskip=-.8em,belowskip=-.7em]
GT/Model: _transfer_block(src, dst.res4, ['4a'] + ['4b%d' % i for i in range(1, (@\HL@)24(@\HLoff@))])
\end{lstlisting}
& 
chainer
&
Y
\\  \midrule

11 &
\begin{lstlisting}[aboveskip=-.8em,belowskip=-.7em]
return 0.5*(@\HL@)(ts-bs)(@\HLoff@)
\end{lstlisting}
&
\begin{lstlisting}[aboveskip=-.8em,belowskip=-.7em]
GT: return 0.5 (@\HL@)- bs(@\HLoff@)
Model: return 0.5*(ts+bs)
\end{lstlisting}
& 
3b1b/manim
&
N
\\  \midrule
12 &
\begin{lstlisting}[aboveskip=-.8em,belowskip=-.7em]
self.stdinlogOpen = (@\HL@)True(@\HLoff@)
\end{lstlisting}
&
\begin{lstlisting}[aboveskip=-.8em,belowskip=-.7em]
GT/Model: self.stdinlogOpen = (@\HL@)False(@\HLoff@)
\end{lstlisting}
& 
cowrie
&
Y
\\  \midrule
13 &
\begin{lstlisting}[aboveskip=-.8em,belowskip=-.7em]
self.terminal.(@\HL@)stdinlog_open(@\HLoff@)=True
\end{lstlisting}
&
\begin{lstlisting}[aboveskip=-.8em,belowskip=-.7em]
GT/Model: self.terminal.(@\HL@)stdinlogOpen(@\HLoff@)=True
\end{lstlisting}
& 
cowrie
&
Y
\\  \midrule

14 &
\begin{lstlisting}[aboveskip=-.8em,belowskip=-.7em]
progress_total = self.parameter_depth * len(self.http_methods) * (
2 + len(self.integration_response_codes) + len(self.method_response_codes)) (@\HL@)- 1(@\HLoff@)
\end{lstlisting}
&
\begin{lstlisting}[aboveskip=-.8em,belowskip=-.7em]
GT/Model: progress_total = self.parameter_depth * len(self.http_methods) * (
2 + len(self.integration_response_codes) + len(self.method_response_codes))
\end{lstlisting}
& 
Miserlou/Zappa
&
Y
\\  \midrule

15 &
\begin{lstlisting}[aboveskip=-.8em,belowskip=-.7em]
outputs = outputs (@\HL@)*(@\HLoff@) num_units**0.5
\end{lstlisting}
&
\begin{lstlisting}[aboveskip=-.8em,belowskip=-.7em]
GT: outputs = outputs (@\HL@)/(@\HLoff@) num_units**0.5
Model: outputs = outputs * num_units**0.5 / 2 
\end{lstlisting}
& 
Kyubyong/ transformer
&
N
\\  \midrule

16 &
\begin{lstlisting}[aboveskip=-.8em,belowskip=-.7em]
drive_start = drive_letter((@\HL@)path(@\HLoff@))
\end{lstlisting}
&
\begin{lstlisting}[aboveskip=-.8em,belowskip=-.7em]
GT/Model: drive_start = drive_letter((@\HL@)start(@\HLoff@))
\end{lstlisting}
& 
numba
&
Y
\\  \midrule

17 &
\begin{lstlisting}[aboveskip=-.8em,belowskip=-.7em]
value = utils.force_type((@\HL@)gy[0](@\HLoff@).dtype, self.value)
\end{lstlisting}
&
\begin{lstlisting}[aboveskip=-.8em,belowskip=-.7em]
GT: value = utils.force_type((@\HL@)x(@\HLoff@).dtype, self.value)
Model: value = utils.force_type(gy[1].dtype, value)
\end{lstlisting}
& 
chainer
&
N
\\  \midrule

18 &
\begin{lstlisting}[aboveskip=-.8em,belowskip=-.7em]
output = tout.readlines()
\end{lstlisting}
&
\begin{lstlisting}[aboveskip=-.8em,belowskip=-.7em]
GT: output = tout.read().(@\HL@)splitlines()(@\HLoff@)
Model: output = list(tout.readlines()) 
\end{lstlisting}
& 
salt
&
N
\\  \midrule

19 &
\begin{lstlisting}[aboveskip=-.8em,belowskip=-.7em]
join.shared_port = self._ports.(@\HL@)pop()(@\HLoff@)
\end{lstlisting}
&
\begin{lstlisting}[aboveskip=-.8em,belowskip=-.7em]
GT: join.shared_port = (@\HL@)0(@\HLoff@)
Model: join.shared_port = self._ports.pop(0) 
\end{lstlisting}
& 
deepmind/pysc2
&
N
\\  \midrule

20 &
\begin{lstlisting}[aboveskip=-.8em,belowskip=-.7em]
ret['changes']['interface'] = (@\HL@)''(@\HLoff@).join(diff)
\end{lstlisting}
&
\begin{lstlisting}[aboveskip=-.8em,belowskip=-.7em]
GT: ret['changes']['interface'] = (@\HL@)'\n'(@\HLoff@).join(diff)
Model: ret['changes']['interface'] = (' ').join(diff)
\end{lstlisting}
& 
salt
&
N
\\
\bottomrule
\end{tabular}
\end{table*}

\begin{tcolorbox}[colback=blue!5!white,colframe=blue!5!white,arc=0mm,grow to left by=0mm,left=0mm,grow to right by=0mm,left=1.5mm,right=1.5mm,top=1.5mm,bottom=1.5mm]
	\textbf{Finding \arabic{findingCounter}\stepcounter{findingCounter}:}
Throughout the different evaluation setups, \name{} finds the correct fix more often than the code-only baseline. Furthermore, almost all the instances (more than 97\%) fixed by code-only model are also fixed by \name{}
\end{tcolorbox}

\subsection{RQ3: Combining Trace-Guided and Code-Only Learning}


Multitask learning is a deep-learning training setup in which one model learns to perform multiple tasks by training on data encoding multiple tasks. The goal and intuition behind multitask learning is that the model gains a better understanding by looking at the data from different perspectives that reflect the specificity of the task. More formally, the deep learning model trains to find an internal representation or embedding of data that generalizes to both tasks. There has been much work on multitask learning recently, and it has many applications in computer vision, NLP and software analysis.

In this experiment, we want to explore combining the baseline approach with our approach in one model through the multitask learning setup. Therefore, we use the CodeNet dataset for training. The dataset comprises inputs without trace modality (code-only) and inputs with trace modality (our approach). The two types of inputs corresponding to the two tasks are distinguished using two prefixes: \code{<FFT>} (fix from trace) and \code{<FFC>} (fix from code). When querying the multitask model, we first pass a trace-based data point. Then, if the model cannot fix the bug from trace-based input, we pass the equivalent code-only data point. In other words, we ask the model to fix from the trace. Then, if the predicted fixes are incorrect, we ask the model again to fix from code only. Since the multitask model performs twice the UTOPk for each query, we use UTOPk with k in (2, 6, 10, 20) to ensure fairness with other unitask models.

In Table~\ref{tab:multitask}, we compare the UTOPk accuracy of the multitask model with the single-task models. The table shows a slight improvement of the multitask over our default \name{}. From one side, we can understand that training in a multitask setup improves overall performance. Conversely, we can see that code-only and trace-based program repair can be complementary. The approach can benefit from reasoning about the code only when the trace is unavailable or does not help fix the code.

\begin{table}[]
	\caption{Accuracy of a multitask variant of \name{} compared with single-task variants (CodeNet dataset).}
	\label{tab:multitask}
	
	\begin{tabular}{lrrrr}
		\hline
		Models & \multicolumn{1}{l}{UTOP2(\%)} & \multicolumn{1}{l}{UTOP6(\%)} & \multicolumn{1}{l}{UTOP10(\%)} & \multicolumn{1}{l}{UTOP20(\%)} \\ 
		\hline
		Multitask & 71 & \textbf{86} & \textbf{88} & \textbf{90} \\
		\name{} & \textbf{72} & 82 & 87 & 88 \\
		Code-only & 64 & 73 & 78 & 85 \\
		\hline
	\end{tabular}
\end{table}


\begin{tcolorbox}[colback=blue!5!white,colframe=blue!5!white,arc=0mm,grow to left by=0mm,left=0mm,grow to right by=0mm,left=1.5mm,right=1.5mm,top=1.5mm,bottom=1.5mm]
	\textbf{Finding \arabic{findingCounter}\stepcounter{findingCounter}:}
	Training \name{} in a multitask setup, where the model can use code-only mode or trace-based mode, improves over the single-task variant of \name{} and encodes both \name{} and code-only baseline in one model.
\end{tcolorbox}

\subsection{RQ4: Effectiveness Depending on the Kind of Bug}

In this experiment, we want to explore further
how well the approach performs on different kinds of bugs. 
For that, we first categorize the test data sampled from CodeNet data into five categories. Arithmetic bugs are instances where a numeral operand or operator is wrong in an arithmetic expression, e.g., \code{speed = distance * time} instead of \code{speed = distance / time}. The Varmisuse category represents bugs where the wrong variable is used in a statement. For example, row 16 of Table~\ref{tab:realworld found} represents an example of a varmisuse bug.
The category Functions include missing function call on an expression, calling the wrong function, or calling a function incorrectly. For example, in the following code: \code{graph = sorted(graph.items())}, the function sorted is called in the wrong way because it is missing the initialization of the argument \code{key}, and thus the code should be: \code{graph = sorted(graph.items(), key=lambda x:len(x[1]))}. The bugs related to data structures comprise bugs where a data structure is wrongly manipulated, for example, \code{list.push(0)} instead of \code{list.pop(0)}. Finally, control flow bugs occur in control flow statements such as \code{if, for} and \code{while}. Each category has 20 instances. The second step in our experiment is to use the model trained on the CodeNet dataset to evaluate each category. Similar to previous experiments, we calculate UTOP k for each category where k is 1, 3, 5, or 10. 

Table~\ref{tab:bugstypes} summarizes the model's performance across the five categories. The model performance on different categories is close to the overall performance of \name{} except for the Functions category. The drop in performance on the Functions category is mainly because of the trace's lack of guidance to perform the proper fix. In the previous example, adding the \code{key} argument will change the sorting key of the elements of a list or iterable. The sorting is based on the values of the second element of each tuple in a list of tuples. In the end, the difference between the divergence state of the code and the desired state is the order of elements in the list which is difficult to translate into the addition of the specific new tokens: \code{key=lambda x:len(x[1])}.

\begin{table}[]
	\caption{Accuracy of the approach on different bug types.}
	\label{tab:bugstypes}
	\begin{tabular}{lrrrr}
		\hline
		Bug type & \multicolumn{1}{l}{UTOP1} & \multicolumn{1}{l}{UTOP3} & \multicolumn{1}{l}{UTOP5} & \multicolumn{1}{l}{UTOP10} \\ 
		\hline
		Arithmetic & 75 & \textbf{85} & \textbf{90} & \textbf{90} \\ 
		Varmisuse & 75 & \textbf{85} & \textbf{90} & \textbf{90} \\
		Functions & 35 & 55 & 60 & 65 \\
		Data structures & \textbf{80} & 80 & 80 & 80 \\ 
		Control flow & 60 & 80 & 80 & 80 \\
		\hline
	\end{tabular}
\end{table}



\begin{tcolorbox}[colback=blue!5!white,colframe=blue!5!white,arc=0mm,grow to left by=0mm,left=0mm,grow to right by=0mm,left=1.5mm,right=1.5mm,top=1.5mm,bottom=1.5mm]
	\textbf{Finding \arabic{findingCounter}\stepcounter{findingCounter}:}
	Our approach is effective across different kinds of bugs. It is still challenging for \name{} to fix bugs where the edit requires out-of-vocabulary tokens or needs documentation on how to use a specific function or API. 
\end{tcolorbox}





\subsection{Threats to Validity}

\paragraph{\textbf{Generalization to other languages}} Our approach by design is independent of the target programming language. However, our implementation is entirely in Python and for Python programs. We targeted Python because there needs to be more work on Program repair that evaluates on Python code. While our approach is unrelated to the specifications of any language, it is still interesting to evaluate on other languages to investigate whether the same results can be replicated.

\paragraph{\textbf{Bug localization and tracing}} Our approach works under the hypothesis that the developer can localize the buggy line and give the desired state at the point of divergence through debugging or other tools. While there is work on localizing bugs, providing the desired state is still an issue. It might take much work for the developer to accurately provide the desired state in complicated programs or programs manipulating large data structures.


\section{Related Work}
\label{sec:related}

\paragraph{\textbf{Traditional approaches to program repair.}}
Program repair and fixing has been extensively studied in the literature~\cite{goues2019automated, gazzola2018automatic, monperrus2018automatic}.
Traditional program repair often follows a pipeline of steps.
It starts with fault localization~\cite{lou2019can, saha2013improving, zhang2017boosting, yang2020lightweight, kim2019precise, wong2016survey}, followed by searching fix candidates based on various search strategies~\cite{wen2018context, jiang2018shaping, weimer2013leveraging}.
The fixes are then checked against correctness criteria (e.g., executing tests).
Generating fix candidates is often the most expensive part as it requires searching in a combinatorial space of the program fixes.
While there exist strategies to reduce the search space by limiting the fix templates, i.e., using only insertions, deletions, or existing code snippets appeared in the program~\cite{harman2010automated}, but it may not be expressive enough to represent the correct fixes.
\name{} automates the search process \emph{over the entire vocabulary of the program space} by producing fixes directly via an efficient inference pass (using GPUs).

\paragraph{\textbf{Learning-based program repair.}}
Machine learning has been widely used to localize bugs~\cite{lou2021boosting, kim2019precise, li2021fault, qi2021dreamloc, allamanis2017learning} and generate program fixes~\cite{svyatkovskiy2021mergebert, chakraborty2020codit, qi2021dreamloc, dinella2021deepmerge, zhu2021syntax, allamanis2014learning, huq2022review4repair, ding2020patching, zhang2022coditt5}, the two critical steps in automated program repair.
In particular, to localize bugs, Lou et al.~\cite{lou2021boosting} used tests \emph{coverage} and employed a gated graph neural network on ASTs of buggy code and test cases.
DeepBugs~\cite{pradel2018deepbugs} embeds single expression to predict 3 types of bugs: swapped function arguments, incorrect binary operator, and incorrect operand in a binary operation.
Li et al.~\cite{li2021fault} learns a convolution network based on the code coverage, statement dependency, and stack traces to predict bug locations. 
Qi et al.~\cite{qi2021dreamloc} exploits bug reports to localize bugs. 
To generate repairs, Devlin et al.~\cite{devlin2017semantic} and Vasic et al.~\cite{vasic2019neural} leverage a pointer network to predict the fixes for a restricted set of bug classes.
DeepFix~\cite{gupta2017deepfix} learns an attention-based sequence-to-sequence network to localize and generate fixes for C programs.
Codit~\cite{chakraborty2020codit} learns a tree-based network to better capture the structural changes for program fix. 
CoditT5~\cite{zhang2022coditt5} pretrains a large language model to explicitly learn the general edit operations, and finetune it on three downstream code editing tasks.
Unfortunately, none of the existing program repair approaches consider dynamic program traces, which provide an important hint to repair programs.
As a result, the class of bugs that \name{} can repair is much broader than prior works.

\paragraph{\textbf{Learning representations for programming tasks.}}
With abundant open-source software and computing power available, deep neural networks have been broadly used to automate software development processes ~\cite{NeuralSoftwareAnalysis}.
Examples have these applications include 
type inference~\cite{hellendoorn2018deep, pradel2020typewriter, pei2021stateformer, xu2016python}, 
program generation~\cite{ciniselli2021empirical, chakraborty2022natgen, chen2021evaluating, li2022competition}, 
code summarization~\cite{shia2022evaluation, wang2019learning}, 
bug/vulnerability detection~\cite{chakraborty2021deep, ding2022velvet, ding2022towards, patra2021nalin}, 
and code clone detection~\cite{ding2022towards, mehrotra2021modeling, pei2020trex}.
It has been shown that incorporating traces benefits learning program representations for downstream applications that relies on understanding program semantics~\cite{pei2021stateformer, nye2021show, wang2020blended, patra2021nalin, jin2022symlm}.
We demonstrate that feeding the trace divergence to neural networks can significantly improve its effectiveness of repairing program bugs (\ref{subsec:RQ1}).


\section{Conclusions}
In this paper, we present \name{}, a new approach for learning-based program repair that incorporates the execution trace into the input of the deep learning model. The evaluation on multiple datasets shows that \name{} is effective at fixing bugs, improves over code-only baseline approaches, and allows to repair more real bugs. 

\bibliographystyle{ACM-Reference-Format}
\bibliography{referencesMP,referencesMore,referencesRelated}

\end{document}